\documentclass[12pt]{article}
\usepackage[latin9]{inputenc}
\usepackage[a4paper]{geometry}
\usepackage{float}
\usepackage{amsmath}
\usepackage{amssymb}
\usepackage{graphicx}
\usepackage{setspace}
\usepackage[authoryear]{natbib}
\usepackage[unicode=true,
 bookmarks=false,
 breaklinks=false,pdfborder={0 0 1},backref=false,colorlinks=false]
 {hyperref}
\hypersetup{
 colorlinks,linkcolor={dark-red},citecolor={dark-red},urlcolor={dark-red}}
\providecommand{\tabularnewline}{\\}

\usepackage{tikz}
\usetikzlibrary{patterns,decorations}

\newtheorem{res}{Result}

\usepackage{xcolor}
\definecolor{dark-red}{rgb}{0.4,0.15,0.15}
\definecolor{dark-blue}{rgb}{0.15,0.15,0.4}
\definecolor{medium-blue}{rgb}{0,0,0.5}

\begin{document}

\title{Randomization advice and ambiguity aversion\thanks{We thank Michael Greinecker, John Nachbar, Paulo Natenzon, Andrea
Prat, Todd Sarver, Tomasz Strzalecki, Peter Wakker and Jonathan Weinstein for insightful
comments. We are very grateful to the editor, Kip Viscusi, and an anonymous referee who have pushed us to clarify the nature of the results in this paper. The authors gratefully acknowledge the funding from the Weidenbaum Center on the Economy, Government, and Public Policy at
Washington University in St. Louis.}}

\author{Christoph Kuzmics\thanks{University of Graz, Austria, christoph.kuzmics@uni-graz.at}
\and Brian W. Rogers\thanks{Washington University in St. Louis, U.S.A., brogers@wustl.edu}
\and Xiannong Zhang\thanks{Washington University in St. Louis, 1 Brookings
Dr. St. Louis, Missouri, U.S.A. (63130), zhangxiannong@wustl.edu}}

\maketitle
\bigskip{}

\begin{abstract}
We design and implement lab experiments to evaluate the normative appeal of behavior arising from models of ambiguity-averse preferences. We report two main empirical findings. First, we demonstrate that behavior reflects an incomplete understanding of the problem, providing evidence that subjects do not act on the basis of preferences alone. Second, additional clarification of the decision making environment pushes subjects' choices in the direction of ambiguity aversion models, regardless of whether or not the choices are also consistent with subjective expected utility, supporting the position that subjects find such behavior normatively appealing.
\end{abstract}


\newpage{}

\section{Introduction}

Many economic decisions are made under uncertainty that cannot be readily quantified by objective probabilities. Consider saving decisions, that is investing money into bonds or stocks in the presence of inflation and general uncertainties about the economic future. Even in the absence of wars and pandemics, most people find it hard to attach probabilities to the relevant possible events, let alone agree on such an assessment.

The classical paradigm of rational decision making under such uncertainty is subjective expected utility (SEU), underpinned by the axiomatic foundation of \citet{savage54}. The experimental designs of \citet{ellsberg61}, later implemented in many studies, see e.g., the survey of \citet{trautmann2015ambiguity}, have challenged the SEU paradigm. This challenge was mostly on positive, that is, empirical, grounds in that these experiments found that many people behave in a way that is inconsistent with subjective expected utility. The fact, however, that this behavior has been so robust in lab experiments and that many researchers have developed axiomatic foundations for alternative models of decision making that admit such behavior, see references below, may be received as posing a normative challenge to SEU in postulating that a broader set of preference models could be considered normatively appealing.

In this paper we aim to test the normative appeal of these alternative models of decision making under uncertainty. We use the term ``normative'' in the sense of Ellsberg's 1962 PhD thesis, see \citet[pages 22-26]{ellsberg2001risk}, and as in the ``subjective'' definition of rationality given by \citet[p. 5]{gilboa2012rational}: We consider a decision normatively appealing (to the decision maker) if the decision maker (still) makes this choice after thorough reflection. In all of our treatments, subjects are provided with a complete, and fairly standard, description of all payoff-relevant aspects of the environment. We operationalize this reflection in the lab by providing subjects with supplementary descriptions that, while not payoff-relevant, emphasize certain ways to think about the environment. The descriptions are provided in the form of short videos that subjects watch before the elicitation of their choice. The experimental design has been pre-registered at the AEA RCT Registry, see \cite*{KRZregistration}.\footnote{The pre-registered experimental design comprises treatments for what are now two papers. Treatments 2 to 11 in \cite*{KRZregistration} are reported here, while Treatment 1 is reported in \cite*{kuzmics2022ellsberg}.}

Our findings are relevant for all models of ambiguity aversion that are monotone. That is, if one act is better than another act in all states, then the inferior act cannot be chosen when the superior act is available. We call such models \emph{classical ambiguity aversion (CAA) models}.\footnote{\label{fn:ambiguitymodels}This category includes most preference models reviewed in the survey of \citet{machina2014ambiguity}, such as the maxmin expected utility model of \citet{gilboa89}, the Choquet expected utility model of \citet{schmeidler89}, the smooth ambiguity model of \citet{klibanoff05}, the variational and multiplier preference models of \citet{maccheroni06} and \citet{hansen2001robust}, confidence function preferences of \citet{chateauneuf2009ambiguity}, uncertainty aversion preferences of \citet{cerreia11}, and the incomplete preference model of \citet{bewley02}.}

The experimental environment is directly inspired by the hedging argument of \citet{raiffa61} in the context of the \citet{ellsberg61} two-color urn experiment. A \emph{risky urn} contains 49 White and 51 Red balls.\footnote{This variation from 50/50 avoids identification complications that can arise from indifference.}  An \emph{ambiguous urn} also contains 100 balls, each of which is either Green or Yellow, but nothing more is known about the composition of the ambiguous urn. The decision-maker (DM) must choose one of three actions, after which the experimenter draws one ball from each urn. Together, these determine the consequence for the DM, which is either ``Win'' or ``Lose'', with Win being strictly preferred to Lose. We call the choices bets for White, Green, and Yellow.  Each bet Wins if the experimenter draws a ball of the corresponding color, and loses otherwise.  Ellsberg's key insight is that Bet White, while commonly chosen, is incompatible with SEU.\footnote{If Bet White is chosen, it must be weakly preferred to Bet Yellow, so that the DM's subjective probability of a yellow ball being drawn from the ambiguous urn must be at most 49\%.  But then the subjective probability of a green ball being drawn from the ambiguous urn must be at least 51\%, so that Bet Green is strictly preferred to Bet White, a contradiction.}

In this context, the \citet{raiffa61} hedge against ambiguity works as follows. The DM flips a fair coin and bets Green if the coin lands on heads and bets Yellow if the coin lands on tails. This randomized action provides an (objective) winning probability of 50\%, regardless of the color of the drawn ball from the ambiguous urn. As this is higher than the 49\% winning probability from betting White, this action strictly dominates the Ellsberg choice.\footnote{\citet{kuzmics2017abraham}, appealing to results in classical statistical decision theory -- in particular the complete class theorem of \citet{wald47annals}, has shown that a DM who can randomize over choices and commit to following the realized prescription of this randomization can never make choices that are inconsistent with SEU and at the same time consistent with CAA in any decision problem. See also \citet{bade2015randomization}, \citet{oechssler2014unintended}, \citet{azrieli2018incentives}, \citet{baillon2022experimental} for similar results and arguments along these lines. One can, in fact, regard ambiguity aversion as a preference for randomization, see e.g., \citet{eichberger96} and \citet{epstein2010ambiguity}.}

In light of the \citet{raiffa61} argument, what are the possible explanations for a  classically ambiguity-averse DM to nonetheless bet on White in this experiment even though it is dominated? First, it is possible, even likely, that designing a random choice does not occur to subjects as an option. Second, even if a subject recognizes such a possibility, it is possible that they cannot, or choose not to, go through the required construction and reasoning that would allow them to see that betting on White is dominated.\footnote{The argument is based on the reduction of compound lotteries and as \citet{halevy07} found many subjects who make Ellsberg choices also fail to reduce compound lotteries. This failure, however, can hardly be seen as normatively appealing - it is simply a mathematical mistake.  \citet{abdellaoui2015experiments} has found a much weaker association between displayed ambiguity aversion and a failure to reduce compound lotteries.} Suppose  that a subject does recognize the possibility and understands the argument. A third possible explanation is that the subject has no access to a suitable randomization device, nor thinks they can simulate one.  So suppose the subject does have a fair coin. The fourth and final explanation is that the subject lacks the ability to commit to the randomized action. Once the coin flip realizes, the subject could revisit their choice, and if they are ambiguity averse they will want to flip the coin again, and again ad infinitum, with one possible outcome that they bet on White in the end after all.

Classical ambiguity aversion models do not allow us to delve into the reasons behind the choice of White in the presence of the \citet{raiffa61} argument, as these preference models are axiomatized for preferences over the space of all pure acts, see e.g., \citet{seo09}, and not over the set of all mixed acts as in \citet{anscombe63}. This means, however, that whether or not a classically ambiguity averse subject who understands the environment may choose to bet on White depends simply on whether the \citet{raiffa61} hedge (including the commitment to its outcome) is provided as a pure choice or not. If it is given as a pure choice the subject cannot choose to bet on White. If it is not given, they can.

Motivated by these considerations, our experimental treatments provide differential supplementary observations that focus on the hedging argument.  All of the supplementary observations are given after a (common to all treatments) standard and complete description of the environment, and before the elicitation of the subject's choice.  If the standard and complete description of the environment suffices to impart a full understanding of the consequences of each possible action, then the treatment effects of the supplementary observations will be null.  Indeed, there is no marginal payoff-relevant information in the supplementary observations.

To the contrary, our first main finding is that the commentary significantly changes behavior. Since the commentary does not affect a subject's underlying preferences, it changes behavior through modifying a subjects' \emph{understanding} of the environment.  This is only possible if the subject's understanding after the standard description was incomplete or erroneous.  We conclude that many subjects indeed have an incomplete or erroneous understanding after being provided with a standard and complete description of the classical Ellsberg two-urn environment, so that directly applying the revealed preference toolkit to such data may not be appropriate.  Instead, a given choice may reflect an incomplete comprehension of the environment, and therefore be viewed as a mistake, or the result of confusion, rather than as a manifestation of the subject's preference.

Our second main finding concerns the direction of the effects of the supplemental observations. In this regard, and to the extent we can infer preferences from the treatments with commentary, our data supports the broader normative appeal of ambiguity aversion models over SEU in the following sense: When the Ellsberg choice (bet White) is compatible with ambiguity averse preferences but not with SEU, the supplemental hedging observations increase the frequency of Ellsberg behavior; when the Ellsberg choice is incompatible with ambiguity aversion (and so also with SEU), the same observations decrease the frequency of Ellsberg behavior.

The paper proceeds as follows: The experimental design is given in Section \ref{sec:design}. The results are given in Section \ref{sec:res}. Section \ref{sec:disc} offers a discussion of these results and Section \ref{sec:lit} provides a brief survey of related literature before Section \ref{sec:con} concludes. Experimental instructions are presented in the Appendix.

\section{Experimental Design} \label{sec:design}

The slight variation of the Ellsberg two-color urn experiment, outlined above, serves as the baseline and control. We then vary the supplemental observations that subjects receive about the decision-making environment. Our treatments differ from the baseline along two dimensions. First, in addition to the standard options of betting on White, Yellow, or Green, in some treatments subjects are presented with an additional fourth option, which executes a bet on either Green or Yellow according to the outcome of a fair coin to be tossed by a third party after the balls are drawn from the urns - the Raiffa hedge. Note that this option also serves as a commitment device for randomization, as the bet will be executed by the experimenter on the subject's behalf.  Recall that the compatibility of the Ellsberg choice (bet White) with classical ambiguity aversion models hinges on the presence or absence of this option.

Second, after the environment is fully described as transparently as possible, in some treatments subjects watch short videos before making their (single) choice. The videos, while all factually correct, emphasize different aspects of the consequences of using the randomization device, in ways that we hypothesize may change some subjects' understanding of the merits of the Raiffa hedge.

In the main treatment, the coin flip bet is included as an option, and subjects are presented with a single video (denoted $V1$ and available \href{https://www.dropbox.com/s/y4yyb81slzcupks/V1p.mp4?dl=0}{here}) containing supplemental observations that describe the hedging argument of \citet{raiffa61}.\footnote{Transcripts of all videos are included in the appendix for an offline audience.} It describes the outcome of betting on the coin conditional on the outcome of the ball drawn from the ambiguous urn. It states that the winning probability using that option is 50\% in either case (green ball or yellow ball drawn), and concludes by reminding the subject that betting on White wins with probability 49\%.

This video, and each one of our other videos, does not explicitly advocate for any particular choice, so that it contains observations rather than advice.  The transcripts of the videos are read by an anonymous (to subjects) third party to avoid a perception that the experimenters are giving implicit advice.

Partly to control for a possible experimenter demand effect, we designed a second video (denoted $V2$ and available \href{https://www.dropbox.com/s/61tevb31hmt0zam/V2p.mp4?dl=0}{here}), in which the structure of the argument and the language is symmetric to the first video. It describes the outcome of betting on the coin conditional on the outcome of the coin flip. It states that no known winning probability can be specified in either case (heads or tails), and concludes by reminding the subject that betting on White wins with probability 49\%. Again, it does not advocate for any particular choice.

We ran treatments utilizing exclusively this second video, as well as treatments in which subjects were presented with both videos before eliciting their choice (in both orders; there were no order effects).\footnote{In sessions using both videos, we showed one video first, and the other video next. We did this in both orders. Then after both videos were shown, the subjects were provided with additional time to revisit any portions of either or both of the two videos before proceeding to enter their bet. During this time subjects could pause, rewind, and switch between videos as they liked. We thereby tried to minimize possible order effects and, indeed, there is no evidence that the order of the videos has any effect, and so we have pooled that data in our analysis.}

As we want to understand the effect of the supplementary observations independently from the effect of presenting the hedging device as an explicit option, we ran a parallel set of treatments with similar videos but where the available options were simply bets for White, Green, or Yellow, as in the baseline case, without the coin flip option. In these treatments, the Ellsberg choice remains compatible with CAA models. We varied the videos slightly to accommodate the different choice set. First, as there was no explicit coin, before showing either $V1$ or $V2$, we showed a preliminary video (denoted $V0$ and available \href{https://www.dropbox.com/s/jrkjm39gxrv94yv/V0.mp4?dl=0}{here}) in which it was explained that a subject could imagine a virtual coin toss, and then bet on Green/Yellow according to the outcome. Second, in videos $V1$ and $V2$ the coin toss was referred to as a virtual coin toss. We refer to the treatment with the explicit hedge/commitment option as ``Coin'' and those without it as ``No Coin'' treatments.

\section{Results} \label{sec:res}

Table \ref{tab:result2} summarizes the main findings.

\begin{table}[H]
\centering{}%
\begin{tabular}{c|ccc||c|cccc}
 & W  & G  & Y  &  & W  & G  & Y  & C\tabularnewline
\hline
Baseline  & 45\%  & 41\%  & 14\%  & Baseline  & 37\%  & 26\%  & 11\%  & 26\%\tabularnewline
 & {\footnotesize{}{}{}{}{}{}{}{}{}{}(26/58)}  & {\footnotesize{}{}{}{}{}{}{}{}{}{}(24/58)}  & {\footnotesize{}{}{}{}{}{}{}{}{}{}(8/58)}  &  & {\footnotesize{}{}{}{}{}{}{}{}{}{}(20/54)}  & {\footnotesize{}{}{}{}{}{}{}{}{}{}(14/54)}  & {\footnotesize{}{}{}{}{}{}{}{}{}{}(6/54)}  & {\footnotesize{}{}{}{}{}{}{}{}{}{}(14/54)}\tabularnewline
\hline
V1  & 27\%  & 55\%  & 18\%  & V1  & 29\%  & 2\%  & 2\%  & 67\%\tabularnewline
 & {\footnotesize{}{}{}{}{}{}{}{}{}{}(12/44)}  & {\footnotesize{}{}{}{}{}{}{}{}{}{}(24/44)}  & {\footnotesize{}{}{}{}{}{}{}{}{}{}(8/44)}  &  & {\footnotesize{}{}{}{}{}{}{}{}{}{}(14/48)}  & {\footnotesize{}{}{}{}{}{}{}{}{}{}(1/48)}  & {\footnotesize{}{}{}{}{}{}{}{}{}{}(1/48)}  & {\footnotesize{}{}{}{}{}{}{}{}{}{}(32/48)}\tabularnewline
\hline
V2  & -  & -  & -  & V2  & 41\%  & 9\%  & 9\%  & 41\%\tabularnewline
 &  &  &  &  & {\footnotesize{}{}{}{}{}{}{}{}{}{}(18/44)}  & {\footnotesize{}{}{}{}{}{}{}{}{}{}(4/44)}  & {\footnotesize{}{}{}{}{}{}{}{}{}{}(4/44)}  & {\footnotesize{}{}{}{}{}{}{}{}{}{}(18/44)}\tabularnewline
\hline
V1+V2  & 58\%  & 28\%  & 13\%  & V1+V2  & 23\%  & 13\%  & 9\%  & 55\%\tabularnewline
 & {\footnotesize{}{}{}{}{}{}{}{}{}{}(35/60)}  & {\footnotesize{}{}{}{}{}{}{}{}{}{}(17/60)}  & {\footnotesize{}{}{}{}{}{}{}{}{}{}(8/60)}  &  & {\footnotesize{}{}{}{}{}{}{}{}{}{}(13/56)}  & {\footnotesize{}{}{}{}{}{}{}{}{}{}(7/56)}  & {\footnotesize{}{}{}{}{}{}{}{}{}{}(5/56)}  & {\footnotesize{}{}{}{}{}{}{}{}{}{}(31/56)}\tabularnewline
\end{tabular}\caption{\label{tab:result2} Summary of data (left: without a randomization device; right: explicit randomization option).}
\end{table}

We summarize the key findings from this data in the following three results.

\begin{res} \label{res:preferences} If preferences alone dictate choices, the supplementary observations contained in the videos should have no effect on behavior. For the No Coin (Coin, resp.) treatments, pooling the data for Green and Yellow (being conservative), the p-value for the null hypothesis that choice frequencies are the same in the baseline and the $V1$ video treatment is $0.067$ ($ < 0.001$, resp.)\footnote{All statistical tests performed in this paper are likelihood ratio tests.}, and for the null hypothesis that choice frequencies are the same in the baseline and the neutral $V1+V2$ video treatment is $0.142$ ($0.007$, resp.).
\end{res}

\begin{res} \label{res:lessambiguity} The p-value for the null hypothesis that neutral video observations ($V1+V2$) does not decrease the choice of White relative to the baseline is $\ge0.5$ for the No Coin treatments, and $0.057$ for the Coin treatments.
\end{res}

\begin{res}\label{res:commentary}\footnote{Readers should be aware that we initially planned only to test the two null hypotheses given in Results \ref{res:preferences} and \ref{res:lessambiguity}. The reported test provided here is an afterthought, after we have already seen the data, and is meant only to communicate that there is a significant (perhaps not meant in the statistical sense) and interesting difference of the effect of video commentary between the Coin and No Coin treatments.}
Without supplemental observations there is no significant difference in the frequency of Bet White between the Coin and No Coin treatment, p-value $0.402$. With supplemental observations ($V1 + V2$) there is a significant difference in the frequency of Bet White between the Coin and No Coin treatment, p-value $< 0.001$.
\end{res}

\section{Discussion} \label{sec:disc}

There is firm evidence that behavior across treatments is not a pure consequence of underlying preferences combined with a complete understanding of the environment. The observations contained in the videos cannot change preferences as they do not change any of the payoff-relevant considerations. Rather, any differences in behavior must come from differences in the subjects' understanding.

Suppose we adopt the view that choices after studying both videos indeed reveal preferences, since subjects may have a more complete understanding of the environment and the consequences of their choices after considering the observations contained therein. Even then, 23\% of subjects, i.e., those who chose Bet White in the Coin treatment, have preferences different from any CAA model. The remaining 77\% of subjects make choices in the Coin treatment that can be explained by CAA as well as SEU.

However, the No Coin treatments provide an interesting contrast. In these treatments, Bet White is undominated and the supplemental observations \emph{increase} the frequency of Bet White relative to the baseline description. If, again, we view the choices after studying both videos as revealing preferences, the choice of Bet White made by 58\% of subjects is inconsistent with SEU but is consistent with CAA models.

Together, these findings suggest that CAA models have broader normative appeal than (the narrower theory of) SEU despite their descriptive problems in some environments. Of note, we asked subjects (in a non-incentivized post-experiment questionnaire) if their preference became more or less clear after watching the videos. In the No Coin treatments 17\% (27 out of 162) of subjects reported that their preferences became ``less clear'' after watching the videos, which is much higher than the 3\% (6 out of 202) reporting the same in the Coin treatment ($p<0.001$), calling into question the presumption that behavior directly reveals preferences, especially in the No Coin treatments. One interpretation is that many subjects find monotonicity to be a normatively appealing property, yet lack the sophistication to identify its consequences.

We conclude this discussion by considering preferences that may depend on the timing of the resolution of uncertainty, as in the models of \citet{seo09}, \citet{saito15}, and \citet{ke2020randomization}. Such models are not classical, as they are not monotone.\footnote{Motivated by the state separability embedded in monotonicity, \citet{bommier2017dual} provides a model where monotonicity is relaxed.} In the Coin treatments subjects were (truthfully) told, as part of the baseline description of the environment, that the coin flip would be executed after the balls were drawn from the urns and revealed. Thus, the choice of Bet White in the Coin treatments (37\% in the baseline and 23\% after both videos) is even inconsistent with these more flexible models.

Roughly, one could categorize our subjects as follows. There is one group of subjects ($\ge$ 23\%) who make choices inconsistent with all models of ambiguity aversion. There is a second group of subjects ($\le$ 58\% - 23\% = 35\%) who make choices consistent with ambiguity aversion but not with SEU. The remainder make choices consistent with SEU. Subjects in the second group would have a demand for randomization devices as they cannot, to their satisfaction, create and commit to randomized choices themselves.

\section{Related Literature}

\label{sec:lit}

A number of papers have studied the consistency of subjects' choices across decision problems. These include \citet{binmore2012much}, \citet{stahl2014heterogeneity}, \citet{voorhoeve2016ambiguity} and \citet{crockett2019ellsberg}. This literature finds, on the whole, that relatively few subjects make consistent choices, and those who do tend to be ambiguity-neutral. The lack of consistency can be interpreted as evidence against people choosing according to a clear preference. However, ambiguity-averse DMs may find inconsistent choices to be a useful hedge against ambiguity, see e.g., \citet{kuzmics2017abraham} and \citet{azrieli2018incentives} for more general arguments.\footnote{This problem persists under many different preference elicitation schemes. See also e.g., \citet{bade2015randomization}, \citet{oechssler2015test}, \citet{baillon2022experimental}, which builds on earlier work on eliciting non-expected utility preferences under objective uncertainty by e.g., \citet{Karni87}.} Our single choice design is immune to such problems. This is why we constrained our design to a single incentivized elicitation per subject, even at the cost of forgoing the ability to conduct within-subject analyses across treatments.

\citet{spears2009preference}, \citet{dominiak2011attitudes}, and \citet{oechssler2016hedging} study experiments in which subjects are given the Raiffa hedge as an option, similarly to our baseline Coin treatment (without supplemental commentary). Generally, they find very few subjects choosing this option, with more subjects instead choosing a dominated option. This too is evidence against CAA models. They also find that subjects do not care about the timing of the resolution of uncertainty, evidence even against the non-CAA models of \citet{seo09}, \citet{saito15}, and \citet{ke2020randomization}. Relatedly, \citet{baillon2022randomize} study how subjects respond to two different forms of randomized incentive mechanisms to elicit ambiguity aversion preferences. Somewhat in contrast to the above mentioned literature, they find that, while about 50\% of subjects make an Ellsberg-like choice in a treatment without randomization (akin to a treatment without a coin toss), only 25 to 29\% of subjects make this choice when there is randomization (akin to a treatment with a coin toss). Interestingly, however, and consistent with the above mentioned literature, subjects do again not seem to care about the timing of the resolution of uncertainty.

We add to this literature by focusing on the possible effects on subjects' choices of providing explicit descriptions of the Raiffa hedge in the form of video commentaries. We thus add to these findings that such observations significantly influence behavior, and does so in directions that support the appeal of CAA models.

\citet{nielsen2022choices} study the normative appeal of key axioms of rational decision making under risk. They ask subjects two things: whether they would agree that their choices should satisfy a certain axiom, and to make a choice from a set of lotteries. For all those subjects whose two answers are inconsistent, they point this out to them and ask them if they would reconsider their answers. They then find that of all those subjects whose answers are inconsistent, 47\% revise their choice to be consistent with the axiom, while 13\% revise their wish to be consistent with the axiom. Our approach has a similarities with that of \citet{nielsen2022choices}. We also show subjects an axiom, if one can call it that, by giving them video commentary about the consequences of the \citet{raiffa61} hedge, i.e., of making choices based on a coin flip. We also have a similar goal as in \citet{nielsen2022choices}: we are interested if subjects might make choices by mistake because they do not fully understand the consequences of their choices. In contrast to \citet{nielsen2022choices}, however, and this is the main difference, we here do not study decision making under risk, but under ambiguity. Our ``axiom'' of interest, the \citet{raiffa61} hedge, is, therefore, quite different. We think of the \citet{raiffa61} hedge not so much as an axiom, but more of an argument. Therefore, we do not just ask subjects if they agree with the argument, but provide two neutrally phrased commentaries about the consequences of the coin toss, one implicitly advocating for and one against the \citet{raiffa61} hedge. Some of our findings, while in a different domain of choice problems, are, nevertheless consistent with those of \citet{nielsen2022choices}: in both domains of decision making under risk \citep{nielsen2022choices} and decision making under ambiguity (this paper) there is evidence of people making choices that they themselves see as mistakes after induced contemplation. We then add to this finding evidence in which way these mistakes are made and in which way they are corrected for the domain of decision making under ambiguity.

Finally, several studies test, in various other ways, the normative appeal of ambiguity aversion preference models.\footnote{\citet{alnajjar09} provide normative arguments against ambiguity aversion.} The closest to our design is that of \citet{slovic1974accepts}, who give subjects written advice for and against \citet{allais53} and \citet{ellsberg61} choices. However, their advice is built around the independence axiom, and so concerns a quite distinct domain. \citet{jabarian2022two} also study the effect of a form of advice on subjects' decisions in a framework with ambiguity aversion. Their framework involves two independent draws from the same two-color ambiguous urn (and two draws from a 50-50 risky urn) in which many subjects make dominated choices, similarly also to \citet{yang2017testing} and \citet{kuzmics2022ellsberg}. Subjects win if they draw two balls of the same color from the urn that they choose, making betting on the ambiguous urn a (weakly) dominant choice - as the more extreme the ball distribution in the ambiguous urn the higher the chance of drawing two balls of the same color. \citet{jabarian2022two} have treatments in which subjects are given additional decision problems that should help them understand the mechanism why a choice is dominated. They find that while subjects do seem to understand the mechanism, they, nevertheless, do not seem to transfer this knowledge to the original problem in which they make dominated choices regardless. Finally, \citet{keller2007examination}, \citet{trautmann2008causes}, \citet{charness2013ambiguity}, and \citet{keck2014group} study decision problems with ambiguity in groups (or under peer observation) and find, on the whole, that group discussion and related phenomena tend to lead to more ambiguity-neutral choices.

\section{Conclusion} \label{sec:con}

We have subjected classical preference models of ambiguity aversion models to tests of their normative appeal with experiments that stay close to the original Ellsberg (two-color urn) design.

We find that subjects' choices are affected by payoff-irrelevant commentary. This implies that at least one of the two treatments, without or with commentary, does not allow the full revelation of subjects' preferences.

At least some of our subjects do seem to see a certain normative appeal in the kind of behavior prescribed by classical models of ambiguity aversion and, in particular, the monotonicity axiom. Giving subjects access to additional commentary, in the form of short video clips, results in behavior that is more consistent with these models.

The nature of this normative appeal suggests that people, after sufficient reflection, would have a demand for the ability to commit to randomized choices, a demand which one would surmise should be observable. It would be interesting to identify instruments outside the lab, in the various areas of application of ambiguity aversion models, that could serve to satisfy this demand.

We also find that our subjects lack a complete and perfect understanding of their decision environment and how their choices map into final outcomes, in spite of the fact that we did our best to describe the environment completely and accurately.  If this is the case, then there is room for further descriptions to influence behavior.  We have shown that this is indeed readily observable, using the relatively weak instrument of short video clips providing commentary on the hedging argument of Raiffa.  This means that in classical designs, it may be necessary to view a given choice as arising from a combination of preferences and how the subject understands the environment, where the second channel is non-trivial.  Accordingly, a given choice may not provide direct evidence for or against any given preference model.

\appendix

\section{Experimental Design}

\subsection{Experiment details}

\begin{figure}[H]
\centering{}\includegraphics[scale=0.35]{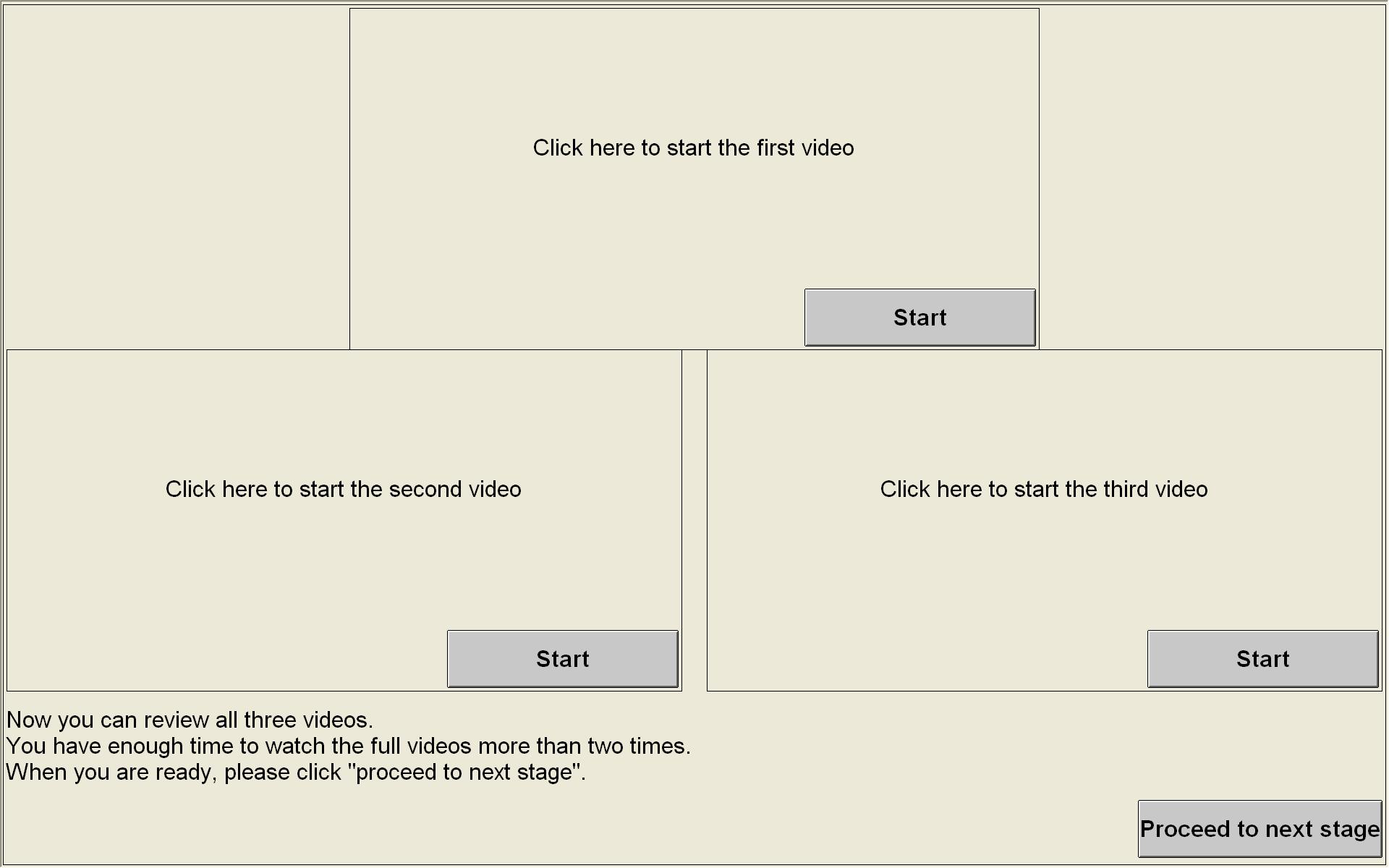}\caption{\label{fig:SS5} Second experiment: Video review. After seeing videos
and before making their choices, subjects had the chance to re-visit
all videos they watched before.}
\end{figure}

The experimental sessions took place in April and May of 2018, and February of 2020. The experiment was conducted at the Experimental and Behavioral Economics Laboratory (EBEL) at University of California, Santa Barbara. There are two waves of data collection. In the first wave, 176 students participated in 10 sessions and the average session length was 60 minutes. In the second wave, 213 students participated in 12 sessions and the average session length was 60 minutes. In all sessions, subjects answered exactly one incentivized question, which was related to guessing the color of a ball. If the guess was correct, the subject received 10 USD, and 0 otherwise. The show-up fee for all sessions was 5 dollars. At the end of each session we conducted a short questionnaire. The questions were not incentivized, but we emphasized that answering these questions would be helpful for our research. The experiment was programmed using z-Tree \citep{fischbacher2007z}. See Figure \ref{fig:SS5} for a screen shot.

\subsection{Physical environment}

\begin{figure}[H]
\centering{}%
\begin{tabular}{cc}
\includegraphics[scale=0.045]{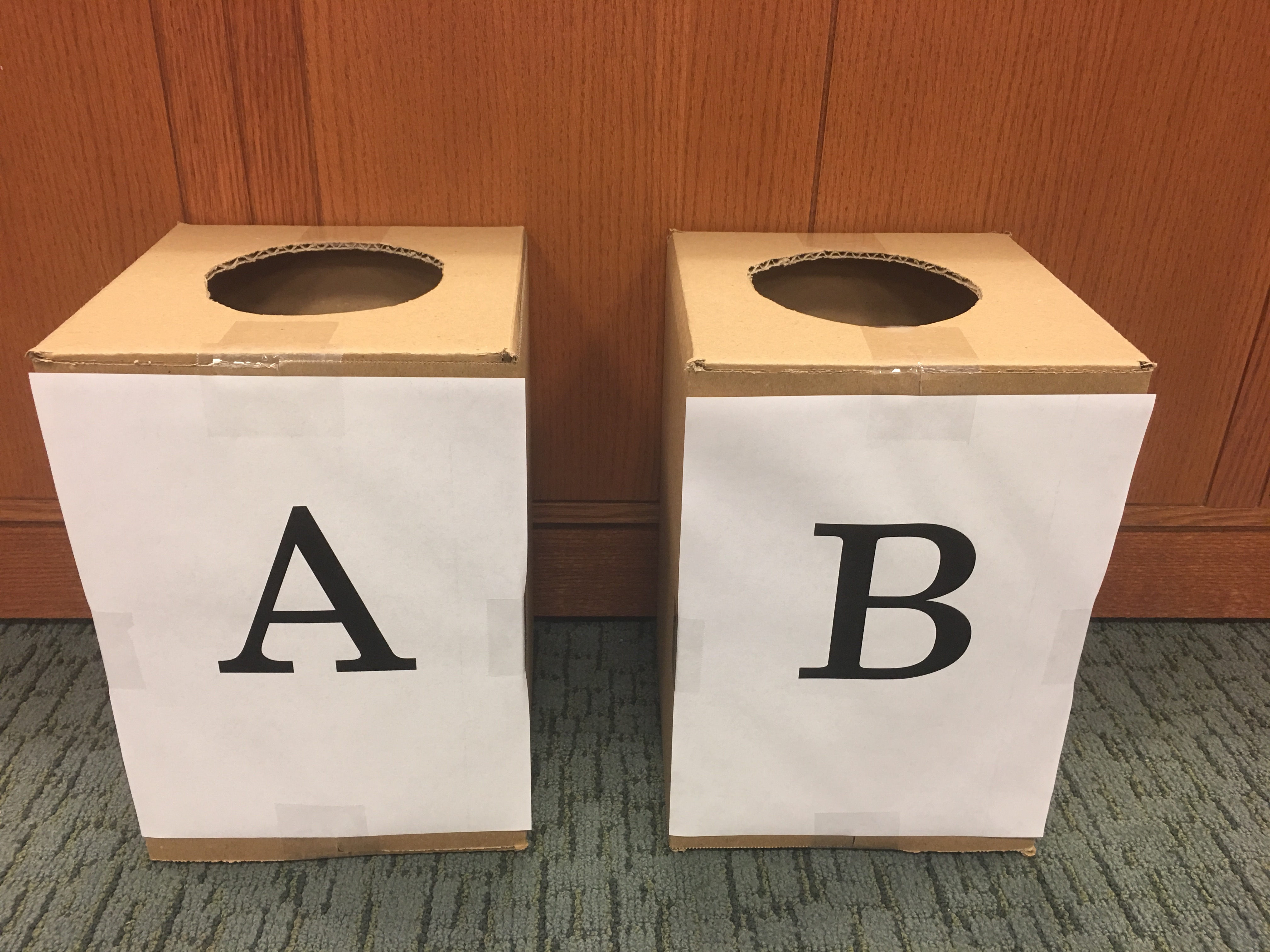}  & \includegraphics[scale=0.045]{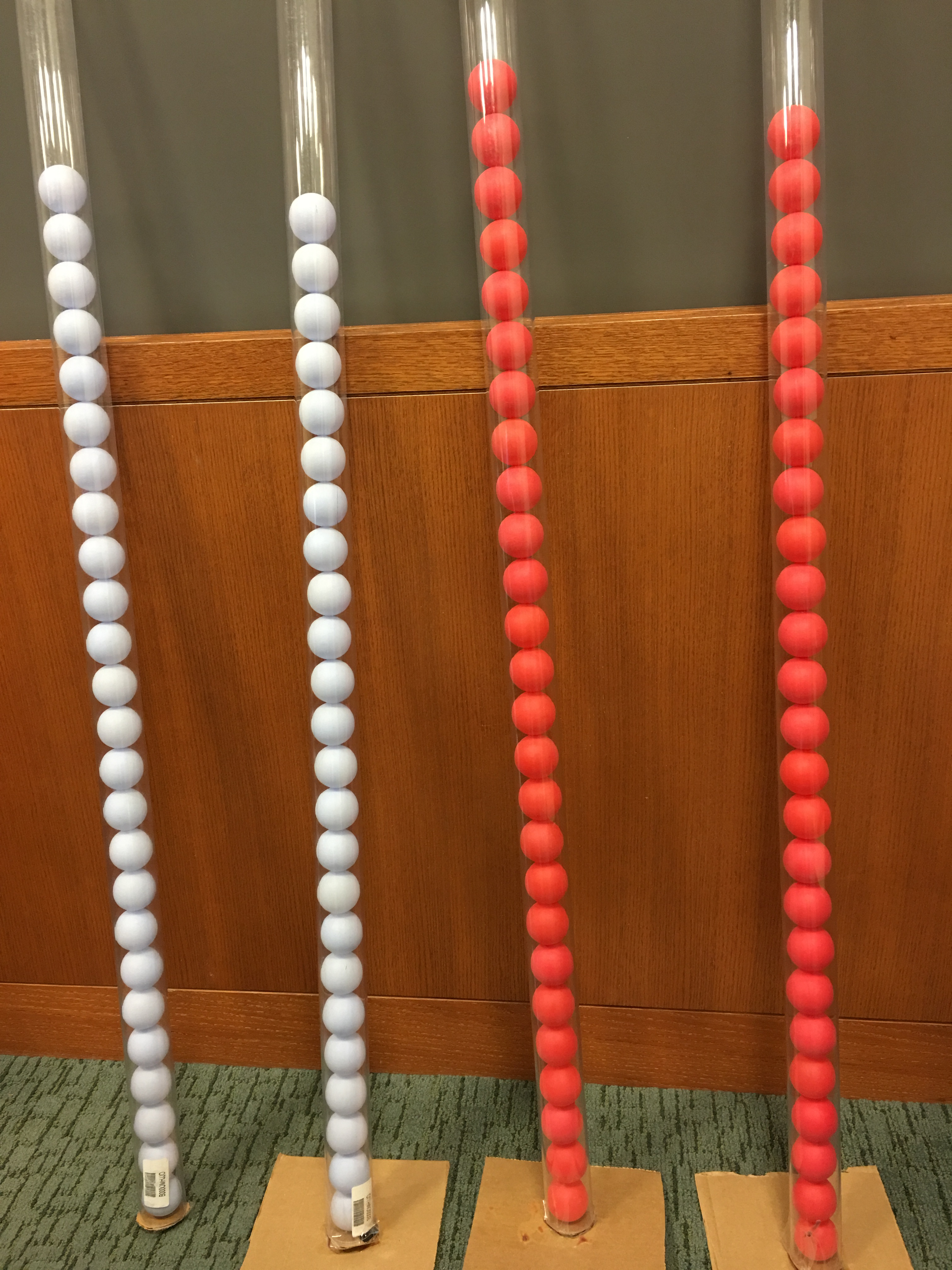}\tabularnewline
a  & b\tabularnewline
\end{tabular}\caption{\label{fig:Boxes}Boxes}
\end{figure}

In all sessions, the urns and states were implemented using two cardboard boxes and colored ping-pong balls. During the experiment (and in what follows), we refer to the two containers as Box A and Box B. A photo of the boxes can be found in Figure \ref{fig:Boxes} (a). The protocols we used were guided by the desire to be as clear and transparent as possible. Box A contained 49 white and 51 red balls. The balls were displayed in clear plastic tubes at the beginning of the experiment so that subjects could easily see that there were two more red than white balls. Photos of the tubes are included as Figure \ref{fig:Boxes} (b). After showing the balls to subjects, they were poured into Box A. On the other hand, it was important that the exact contents of Box B were unknown. We therefore informed subjects that Box B contained 100 balls, each of which was either green or yellow, but we were intentionally not telling them anything further about the contents. Box B was shaken so that it was credible that it contained the same number of balls as Box A. After this presentation, subjects were told that they could inspect all the boxes and balls at the conclusion of the experiment if they so desired.

In each session, one subject was randomly selected to act as a monitor. The monitor was the person who physically conducted all draws of balls and displayed their colors to the other subjects, as well as coin flips, as relevant.

\subsection{A Screen Shot}

\subsection{Questionnaire}

Table \ref{tab:Questionnaire-2} shows the additional questions that we asked at the end of the experiment.

%\begin{centering}
\begin{table}[H]
\begin{tabular}{|c|}
\hline
\textbf{Questions asked in all groups}\tabularnewline
Gender (Male, Female, Prefer not to tell)\tabularnewline
Major\tabularnewline
How many Green balls do you think there are in Box B?\tabularnewline
How many Yellow balls do you think there are in Box B?\tabularnewline
Was there any part of the experiment that was unclear?\tabularnewline
After watching the videos, my preference over choices was: \tabularnewline
(More clear, Less clear, Unchanged, I don't know)\tabularnewline
\hline
\hline
\textbf{Questions asked when V1 and V2 are both presented}\tabularnewline
Which video do you think is more compelling? (V1, V2, Equal)\tabularnewline
\hline
\hline
\textbf{Questions asked in when option Coin is not available}\tabularnewline
Do you find it difficult to simulate a coin toss in you head? (Yes,
No)\tabularnewline
\hline
\hline
\textbf{Questions asked when subjects chose White ball in Box A}\tabularnewline
Why did you choose White?\tabularnewline
\hline
\hline
\textbf{Questions asked when subjects chose Green or Yellow ball in
Box B}\tabularnewline
Why did you choose Green or Yellow?\tabularnewline
\hline
\hline
\textbf{Questions asked in no-video treatments}\tabularnewline
Why did you recommend this to your friend?\tabularnewline
\hline
\end{tabular}%\par\end{centering}
%\begin{centering}
\centering{}\caption{List of questions asked in the questionnaire\label{tab:Questionnaire-2}}
\end{table}

In the battery of sessions for the experiment, different treatments required different questions. The first block lists the questions that we asked in all treatments. The second block lists questions that are asked when both videos are presented. We denote by V1 the video in favor of the hedging argument and by V2 the video with the counter-argument. The third block lists questions that are asked when subjects are offered only three options and must implement the randomization with a virtual coin on their own. The fourth and fifth blocks list questions contingent on subjects' choices. In the treatments in which videos are not shown to the subjects before their incentivized choices, we showed the video before the questionnaire and asked for their ``recommendation'' in the questionnaire. We further asked for their reasoning. This is listed in the last block.

\subsection{Instructions}

We attached the instruction of the most comprehensive treatment. In this treatment, we provided the subjects with 4 options and both videos. Instructions for all the other treatments are written in a similar fashion.
\begin{center}
\textbf{\large{}{}{}{}{}{}{}{}{}{}{}{}{}Instructions}{\large\par}
\par\end{center}

Welcome to the experiment! Please take a seat as directed. Please wait for instructions and do not touch the computer until you are instructed to do so. Please put away and silence all personal belongings, especially your phone. We need your full attention for the entire experiment. Adjust your chair so that you can see the screen in the front of the room. The experiment you will be participating in today is an experiment in decision making. At the end of the experiment you will be paid for your participation in cash. Each of you may earn different amounts. The amount you earn depends on your decisions and on chance. You will be using the computer for the experiment, and all decisions will be made through the computer. DO NOT socialize or talk during the experiment. All instructions and descriptions that you will be given in this experiment are factually accurate. According to the policy of this lab, at no point will we attempt to deceive you in any way. Your payment today will include a \$5 show up fee. One of you will be randomly selected to act as a monitor. The monitor will be paid a fixed amount for the experiment. The monitor will assist us in running the experiment and verifying the procedures. If you have any questions about the description of the experiment, raise your hand and your question will be answered out loud so everyone can hear. We will not answer any questions about how you ``should'' make your choices. As I said before, do not use the computer until you are asked to do so. When it is time to use the computer, please follow the instructions precisely.

We will now explain the experiment. There are two containers on the table that we will refer to as Box A and Box B. This is Box A. The Box is empty. Box A will contain 100 ping pong balls. Each of the balls in Box A will be either White, like this, or Red, like this. Specifically, Box A will contain exactly 49 White balls and 51 Red balls, for a total of 100 balls. You don't have to remember these numbers. When it is time to make a decision, we will remind you of these numbers. We have counted and displayed the balls in these tubes to make it easier to show the contents of Box A. There are 25 white balls in this tube and 24 in this tube, for a total of 49 white balls. There are 25 red balls in this tube and 26 red balls in this tube, for a total of 51. We will now pour these balls into Box A and shake it to mix the balls together. This is Box B. We have already filled Box B with 100 ping pong balls. Each ball is either Green, like this, or Yellow, like this. We will not reveal the exact numbers of Green and Yellow balls. Instead, you know only that there are 100 balls in total, consisting of some combination of Green and Yellow balls. We will now shake Box B to mix the balls up. At the end of the experiment, you will have an opportunity to inspect the Boxes and ping pong balls if you wish. In a few moments we will ask the Monitor to draw one ball from each Box for everyone to observe. You will be asked to choose from several options that correspond to guessing the color of a ball that the Monitor draws. If your guess matches the result, you will receive 10 dollars in additional to the show up fee. If your guess does not match, you will receive 0 dollars in addition to the show up fee.

We will now start the experiment. On the computer desktop you will find a green icon named zleaf. Double click it now. Now there should be a welcome screen. Type your name and click the OK button in the welcome screen. One of you has been randomly selected by the software to serve as the monitor. Please raise your hand if you are the monitor. Could you please click the OK button on your screen and come to the front? Now your screen should have changed to ``Please listen to the instructions\textquotedblright. Please leave it like that and do not click OK. In a few moments the Monitor is going to draw one ball from Box A and one ball from Box B. We are going to ask you to bet on the outcome of those draws.

Specifically, you will be able to place one of 4 bets. Let me explain three of these options first. You can bet on White (from Box A), Green (from Box B) or Yellow (from Box B). Notice that you cannot bet on Red. If you bet on the White ball from Box A, then your payoff is not related to the draw from Box B. In other words, if the monitor draws a White ball from Box A, you win. If the monitor draws a Red ball, you lose. Similarly, if you choose Green or Yellow, your payoff only depends on the draw from Box B. For the fourth option, your bet will depend on the outcome of a coin flip. The monitor will flip a coin like this. If the coin lands on Heads, then we will set your bet to Green. If the coin lands on tails, we will set your bet to Yellow. To repeat, we will set your bet to either Green or Yellow, depending on the coin flip result. Again, you don't have to write this down, since we will remind you about all the options when it is time to make your choice.

Before you make you decision, we are going to provide you with some comments on the experiment contained in 2 short videos. The videos are in total about 5 minutes long. After the videos, you will make your choice by selecting one of the four options. We will proceed like this: You will first watch the two videos. Then, you will have a chance to review the videos if you like. During the review session, you can pause or rewind the videos. There will be enough time to watch both videos more than two times in the review session. After the review session, we are going to ask for your choice. After you enter your decision, please wait for others to finish. There will not be any further instructions until all of you make your decisions. Please follow the instructions on the screen and focus on the videos as much as possible. If you finish early, please remain quiet since others may still be watching. Now please put on the headphones provided at your desk and watch the videos. Once you are ready, please click OK.

The monitor is now going to draw the balls. Please look away and draw a ball from Box A and show it to everyone. The color is {[}REALIZED COLOR{]}. Please put the ball back. We will write down the result on the blackboard. Now please look away and draw a ball from Box B and show it to everyone. The color is {[}REALIZED COLOR{]}. Please put the ball back. We will write down the result on the blackboard. Please toss the coin and announce the result. The result is {[}REALIZED SIDE{]}. Please put the coin down. We will write down the result on the blackboard. Now please return to your seat and enter these results into your computer screen, accompanied by an Experimenter. You can now see the outcome and your earnings on the screen. If you have questions about your payoff, please raise your hand.

We will now conduct a short questionnaire. Please wait for the questionnaire to start. The monitor doesn't have to fill the questionnaire. Please complete the questionnaire. Please be as specific as you can in your responses. Answering the question is helpful to our research, but your responses are entirely voluntary. After you finish, please wait for others. We will call you to the front by your participant ID to be paid before leaving. Thank you very much for your participation. This concludes the experiment. We will now begin calling you to the front to be paid before leaving.

\subsection{Video scripts}

\subsubsection{Names and notations}

Recall that we denote the video that explains the Raiffa hedging argument, used in our main treatment, by V1 and its counter argument by V2. In the treatments without an explicit coin flip option, the instructions do not describe a coin. Instead, we show a short video introducing the hedging idea through the use of an ``imaginary coin.'' We call this video V0. V0 is neither in favor of hedging nor against hedging. It merely states the idea of conditioning one's bet on the outcome of a virtual coin flip. We then slightly modified V1 and V2 by changing ``the coin flip option'' to ``the imaginary coin.'' For more details, please see the script below.

To summarize, we have in total five distinct videos, listed in the table below, where ``p'' stands for ``physical coin'' and ``v'' for ``virtual coin.'' \medskip

\begin{tabular}{c|ccccc}
 & V1p  & V2p  & V0  & V1v  & V2v\tabularnewline
\hline
In favor of hedging?  & Yes  & No  & n/a  & Yes  & No\tabularnewline
Against hedging?  & No  & Yes  & n/a  & No  & Yes\tabularnewline
Describe hedging using a real coin?  & Yes  & Yes  & n/a  & No  & No\tabularnewline
Describe hedging using a virtual coin?  & No  & No  & n/a  & Yes  & Yes\tabularnewline
Description of a virtual coin?  & n/a  & n/a  & Yes  & n/a  & n/a\tabularnewline
\end{tabular}

\subsubsection{V0 Script}

Recall that your three options are to choose: a White Ball from Box A, a Green Ball from Box B, or a Yellow Ball from Box B. Let me suggest a new method for choosing how to bet. To use this method, you need to create a random event. So, imagine you have a coin and you can flip it. The coin lands on Heads with probability 50\% and on Tails with probability 50\%. Before the toss, you plan to bet on a Green Ball from Box B if the coin lands on Heads, and on a Yellow Ball from Box B if the coin lands on Tails. Using this rule, you will not bet on a White Ball from Box A. To summarize, you first imagine the outcome of the coin flip. Then you choose Green Ball from Box B if the coin lands on Heads and you choose Yellow Ball from Box B if the coin lands on Tails. \href{https://www.dropbox.com/s/jrkjm39gxrv94yv/V0.mp4?dl=0}{Click to view}

\subsubsection{V1p Script}

Recall that Box A contains 49 white balls and 51 red balls. So, you will win with probability 49\% if you choose the ``White Ball from Box A.'' Let me describe the outcome when you choose the ``Coin flip for green/yellow ball.'' Recall that Box B contains an unknown combination of 100 Green and Yellow balls. So when the Monitor draws a ball from Box B there are two possible cases: the ball can either be Green, or it can be Yellow. Suppose the ball happens to be Green. Now, when the monitor flips the coin, it will land either on Heads or on Tails. Each case is equally likely: the probability of Heads is 50\% and the probability of Tails is 50\%. If the coin lands on Heads, you would bet on Green and win. If the coin lands on Tails, you would bet on Yellow and lose. So, what we have observed is that if the ball from Box B happens to be Green, you would win with probability 50\%. Now suppose that the ball from Box B happens to be Yellow. As before, when the monitor flips the coin, it will land either on Heads or on Tails. Each case is equally likely: the probability of Heads is 50\% and the probability of Tails is 50\%. If the coin lands on Heads, you would bet on Green and lose. If the coin lands on Tails, you would bet on Yellow and win. So, what we have observed now is that if the ball from Box B happens to be Yellow, you would again win with probability 50\%. To summarize, if you choose the option ``Coin flip for green/yellow ball'' you will win with probability 50\% whether the ball from Box B is green or yellow. Therefore, it does not matter how many of the balls are green and how many are yellow, since you will win with probability 50\% in either case. By betting instead on a White Ball from Box A, you will win with probability 49\%. \href{https://www.dropbox.com/s/y4yyb81slzcupks/V1p.mp4?dl=0}{Click to view}

\subsubsection{V2p Script}

Recall that Box A contains 49 white balls and 51 red balls. So, you will win with probability 49\% if you choose the ``White Ball from Box A.'' Let me describe the outcome when you choose the ``Coin flip for green/yellow ball.'' If you choose this option, there are two possibilities: when the monitor flips the coin, it will land either on Heads or on Tails. Each case is equally likely: the probability of Heads is 50\% and the probability of Tails is 50\%. Suppose the coin happens to land on Heads. In this case, you would be betting on a Green Ball from Box B. The chance that betting on a Green Ball from Box B wins depends on how many green balls are in Box B. Since you are not told how many green balls are in Box B, your probability of winning is uncertain. So, what we have observed is that if the coin lands on Heads, your probability of winning is uncertain. Now suppose the coin happens to land on Tails. In this case, you would be betting on a Yellow Ball from Box B. The chance that betting on a Yellow Ball from Box B wins depends on how many yellow balls are in Box B. Since you are not told how many yellow balls are in Box B, your probability of winning is again uncertain. \href{https://www.dropbox.com/s/61tevb31hmt0zam/V2p.mp4?dl=0}{Click to view}

\subsubsection{V1v Script}

Recall that Box A contains 49 white balls and 51 red balls. So, you will win with probability 49\% if you choose the ``White Ball from Box A.'' Let me describe the outcome when you choose the method based on the coin flip I described before. Recall that Box B contains an unknown combination of 100 Green and Yellow balls. So when the Monitor draws a ball from Box B there are two possible cases: the ball can either be Green, or it can be Yellow. Suppose the ball happens to be Green. Now, when you imagine flipping the coin, it will land either on Heads or on Tails. Each case is equally likely: the probability of Heads is 50\% and the probability of Tails is 50\%. If the coin lands on Heads, you would bet on Green and win. If the coin lands on Tails, you would bet on Yellow and lose. So, what we have observed is that if the ball from Box B happens to be Green, you would win with probability 50\%. Now suppose that the ball from Box B happens to be Yellow. As before, when you imagine flipping the coin, it will land either on Heads or on Tails. Each case is equally likely: the probability of Heads is 50\% and the probability of Tails is 50\%. If the coin lands on Heads, you would bet on Green and lose. If the coin lands on Tails, you would bet on Yellow and win. So, what we have observed now is that if the ball from Box B happens to be Yellow, you would again win with probability 50\%. To summarize, if you use the method based on the coin flip, you will win with probability 50\% whether the ball from Box B is green or yellow. Therefore, it does not matter how many of the balls are green and how many are yellow, since you will win with probability 50\% in either case. By betting instead on a White Ball from Box A, you have a known probability of winning of 49\%. \href{https://www.dropbox.com/s/cnzat5tj4iqr5vv/V1v.mp4?dl=0}{Click to view}

\subsubsection{V2v Script}

Recall that Box A contains 49 white balls and 51 red balls. So, you will win with probability 49\% if you choose the ``White Ball from Box A.'' Let me describe the outcome when you choose the method based on the coin flip I described before. If you use this method, there are two possibilities: when you imagine flipping the coin, it will land either on Heads or on Tails. Each case is equally likely: the probability of Heads is 50\% and the probability of Tails is 50\%. Suppose the coin happens to land on Heads. In this case, you would be betting on a Green Ball from Box B. The chance that betting on a Green Ball from Box B wins depends on how many green balls are in Box B. Since you are not told how many green balls are in Box B, your probability of winning is uncertain. So, what we have observed is that if your coin lands on Heads, your probability of winning is uncertain. Now suppose your coin happens to land on Tails. In this case, you would be betting on a Yellow Ball from Box B. The chance that betting on a Yellow Ball from Box B wins depends on how many yellow balls are in Box B. Since you are not told how many yellow balls are in Box B, your probability of winning is again uncertain. So, what we have now observed is that if your coin lands on Tails, your probability of winning is also uncertain. To summarize, if you use the method based on the coin flip, your probability of winning is uncertain if the coin lands on Heads and it is also uncertain if the coin lands on Tails. By betting instead on a White Ball from Box A, you have
a known probability of winning of 49\%. \href{https://www.dropbox.com/s/d1wyqfg1asagbdb/V2v.mp4?dl=0}{Click to view}

\bibliographystyle{abbrvnat}
\bibliography{reference}

\end{document}